# Fresnel incoherent correlation holography with single camera shot

A. Vijayakumar,[a,*] Tomas Katkus,[a] Stefan Lundgaard,[a] Denver Linklater,[a,b] Elena P. Ivanova,[b] Soon Hock Ng[a] and Saulius Juodkazis[a,c,d,*]

[a]Center for Micro-Photonics, Faculty of Science, Engineering and Technology, Swinburne University of Technology, Hawthorn, VIC 3122, Australia.
[b]Department of Physics, RMIT, GPO Box 2476, Melbourne VIC 3001, Australia.
[c]Melbourne Centre for Nanofabrication, ANFF, 151 Wellington Road, Clayton, VIC 3168, Australia.
[d]Tokyo Tech World Research Hub Initiative (WRHI), School of Materials and Chemical Technology, Tokyo Institute of Technology, 2-12-1, Ookayama, Meguro-ku, Tokyo 152-8550, Japan.

**Abstract**. Fresnel incoherent correlation holography (FINCH) is a self-interference based super-resolution three-dimensional imaging technique. FINCH in inline configuration requires an active phase modulator and at least three camera shots to reconstruct objects without the twin image and bias terms. In this study, FINCH is realized using a randomly multiplexed bifocal binary Fresnel zone lenses fabricated using electron beam lithography. A modified hologram reconstruction mechanism is presented which introduces the single shot capability in FINCH. A point spread hologram library was recorded using a point object located at different axial locations and an object hologram was recorded. The image of the object at different planes were reconstructed using decorrelation of the object hologram by the point spread hologram library. Application potential including bio-medical optics is discussed.

*correspondence, e-mail: vanand@swin.edu.au, sjuodkazis@swin.edu.au

## 1 Introduction

Fresnel incoherent correlation holography (FINCH) was developed by Joseph Rosen and Gary Brooker in 2007 using the self-interference principle.[1,2] In FINCH, the object wave is split into two using a randomly multiplexed diffractive lens displayed on a SLM and two images were created. An image sensor was located in between the two images and the self-interference hologram was recorded.[1] Holograms corresponding to three phase shifts (0, $2\pi/3$ and $4\pi/3$) introduced to one of the two diffractive lenses were recorded and projected into the complex space and superposed to produce a complex hologram. The image of the object is reconstructed by numerically propagating the complex hologram to one of the two image planes. FINCH is considered advantageous compared to existing incoherent imaging techniques as FINCH is motionless and non-scanning.[3] FINCH went through several upgradations in course of time which converted FINCH into a robust, reliable and super resolution three-dimensional imaging technique as it is today. In the first design of FINCH,[1] the numerical reconstruction generated background noise due to the random multiplexing of lenses. To improve the signal to noise ratio (SNR) a polarization multiplexing scheme was proposed,[4] in which, the super-resolution imaging capability of FINCH was also realized.[5] In the later studies, the fringe visibility was improved by reducing the path length difference between the two interfering beams.[6] The main drawback associated with FINCH was that the hologram recording required at least three camera shots and therefore needed an active device such as a spatial light modulator and consequently could not record faster events. A modified version of FINCH called as Fourier incoherent single channel holography (FISCH) was proposed by Roy Kelner and Joseph Rosen in 2012 which exhibited the same resolution as FINCH but required only a single camera shot.[7] However, the penalty was paid by an increase in the number of optical components and complicated beam alignment procedures. Many solutions were developed later to reduce the number of camera shots.[8-12] One solution involved a micro polarizer array which was used with a camera and a single camera shot was decomposed into four images which were processed using de-mosaicing and interpolation to fill in the missing pixel information.[8] The technique, even though advanced, it suffered from background noise and is unable to reconstruct complicated objects. Another solution involved multiplexed gratings, and the image sensor was shared among four camera shots with four different phase shift values. In other words, the required multiple temporal shots were converted to spatially separated single shot. In Ref.10, a dual focusing lens was implemented but the twin image could not be removed, and the reconstruction was demonstrated only for a simple object such as a point. An off-axis configuration was demonstrated in Ref.11 which is unable to match the high resolution of FINCH as the perfect beam overlap cannot be achieved in off-axis configuration. In Ref. 12, a geometric phase lens has been applied along with a micro polarizer array in the image sensor and the images were reconstructed with an improved signal to noise ratio in comparison to Ref. 8.

In all the above suggested improvements to develop a single shot capability in FINCH resulted in penalties paid in the form of increased number of active and passive optical components and the overall experimental footprint. With the development of fabrication technologies[13,14] and as research focuses more on reducing the size, weight and overall cost of imaging systems, realizing the optical configurations used in Refs.8-12 is often cumbersome. In this study, we propose a randomly multiplexed bifocal diffractive lens as the only optical element of FINCH. In addition, we propose a modification of the reconstruction mechanism with a manual PSF training technique introduced by Joseph Rosen in the series of developments in coded aperture imaging technology.[15-18] The result is a single camera shot FINCH technique with manual PSF training and decorrelation techniques.

## 2 Methodology

The optical configuration of FINCH with a randomly multiplexed bifocal diffractive lens (RMBDL) is shown in Fig. 1. Light from an incoherent source critically illuminates a multiplane or 3D object and the light diffracted from the object is collected by the RMBDL which splits it into two object waves. One of the object waves is collimated, while the other is focused at a distance and the self-interference pattern is recorded at a plane where the two object waves are perfectly overlapped. The self-interference pattern is recorded by an image sensor. In general, three self-interference patterns are recorded with relative phase shifts of $\Phi = 0$, $2\pi/3$ and $4\pi/3$ and superposed to generate a complex hologram which is propagated numerically to reconstruct the object.[1,4,5] In the proposed method, a PSF library is first recorded using a pinhole located at different axial planes and the image of the object at different planes were reconstructed by a cross-correlation between the object hologram and the PSF library.[16]

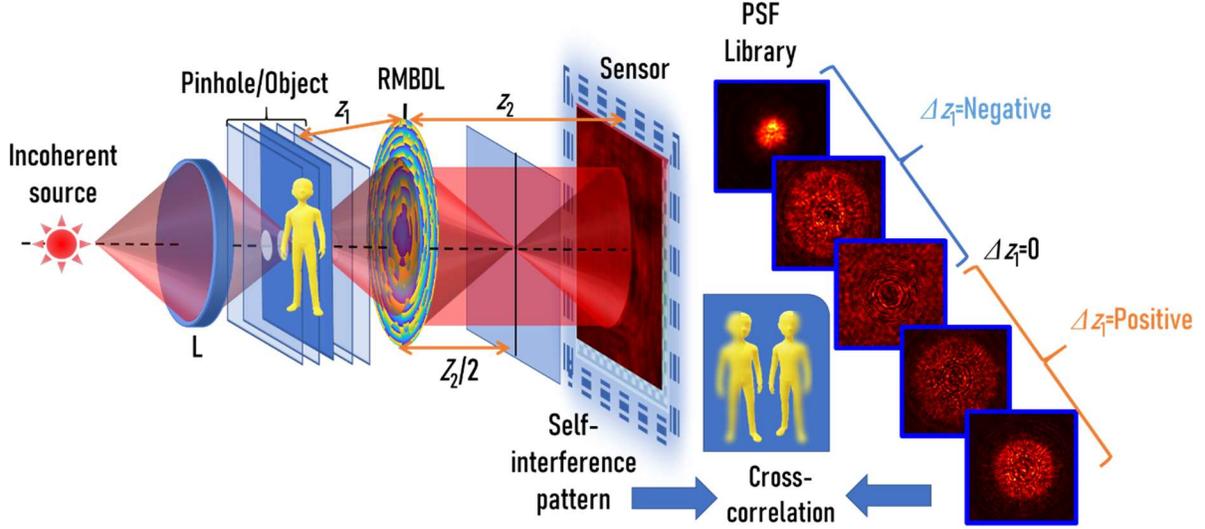

**Fig. 1** Optical configuration of FINCH with a RMBDL and with a modified image reconstruction.

*2.1 Design of RMBDL*

The RMBDL is designed using two Fresnel zone lenses designed for two different configurations. The first Fresnel zone lens (FZL$_1$) is designed for infinite conjugate mode with a focal length $f = z_1$, which is the distance between the object plane and the RMBDL and so the object wave is collimated by FZL$_1$. The second Fresnel zone lens (FZL$_2$) is designed for finite conjugate mode with $u = z_1$ and $v = z_2/2$.[19] The object wave is focused by FZL$_2$ at $z_2/2$ from the RMBDL. The hologram is recorded approximately at $z_2$ from the RMBDL, where the two beams have roughly the same diameter and are perfectly overlapped. The mathematical analysis is carried out from the object plane to the sensor plane. A point located at $(0,0,z_1)$ emits light with an amplitude of $\sqrt{I_o}$ which reaches the RMBDL with a complex amplitude given by $C_1\sqrt{I_o}Q(1/z_1)$, where $C_1$ is a complex constant and $Q(1/z_1) = \exp(j\pi r^2/\lambda z_1)$ is the quadratic phase factor in which $r$ is the radial coordinate. In order to produce a wave with constant phase, the quadratic phase factor must be cancelled by FZL$_1$. Therefore, the complex amplitude of FZL$_1$ must be $\exp(-j\pi r^2/\lambda z_1)$ or the phase of the FZL$_1$ must be $\phi_{FZL1} = -\pi r^2/\lambda z_1$. On the other hand, FZL$_2$ must be designed to convert the complex amplitude $\exp(j\pi r^2/\lambda z_1)$ into $\exp(-j2\pi r^2/\lambda z_2)$ and so the complex amplitude of FZL$_2$ must be $\exp\left\{-j\pi r^2 \lambda^{-1}\left(\frac{2}{z_2} + \frac{1}{z_1}\right)\right\}$. Assuming $z_1 = z_2/2$, the phase of FZL$_2$ can be expressed as $\phi_{FZL2} = -4\pi r^2/\lambda z_2$.

The two FZLs will be randomly multiplexed and fabricated on a glass substrate. Earlier studies indicated that the exclusion of the thickness and refractive index of glass plates introduces substantial spherical aberration into the system resulting in a variation in the focal distances and blurring of the focal spot.[19] Two techniques have been proposed to avoid the spherical aberration. In the first technique, the glass substrate was included in the calculation of the zones of the FZL and in the second case an equivalent and opposite aberration was introduced during fabrication to compensate the spherical aberration. Considering the higher success with the first method, it is adapted for this design. For a thickness $t$ and refractive index $n_g$ of the glass substrate, the phase of the glass substrate is given as $\phi_G = 2\pi n_g t/\lambda$. The phases of the two FZLs after the inclusion of the substrate correction is given as $\phi_{FZL1}' = -\pi r^2/\lambda z_1 - 2\pi n_g t/\lambda$ and $\phi_{FZL2} = -4\pi r^2/\lambda z_2 - 2\pi n_g t/\lambda$. A random phase function $\phi_r$ with a predefined scattering ratio is synthesized using Gerchberg-Saxton algorithm (GSA) and binarized to two levels as $M = \text{round}(\phi_r/2\pi)$. An inverted image of $M$ given as $1 - M$ is synthesized next. The RMBDL is designed by randomly multiplexing the two FZLs using the random phase functions as

$$\phi_{RMBDL} = \{-(4\pi r^2/\lambda z_2) - (2\pi n_g t/\lambda)\}M + \{-(\pi r^2/\lambda z_1) - (2\pi n_g t/\lambda)\}(1 - M). \qquad (1)$$

The RMBDL was designed for $z_1$ = 5 cm, $z_2$ = 10 cm, $\lambda$ = 617 nm, diameter of the FZLs $D$ = 5 mm, $t$ = 1.1 mm, $n_g$ = 1.5 for ITO coated glass plates. The images of the FZLs before and after substrate correction, GSA algorithm for random matrix synthesis and the synthesis of RMBDL with random multiplexing is shown in Fig. 2. A scattering ratio of $\sigma = b/B = 0.1$ was selected and iterated 50 times. The final RMBDL was binarized to two levels for ease of fabrication as shown in Fig. 2. The binarization step may result in the generation of multiple diffraction orders and for analysis in the next section, the contribution in the first diffraction order is only considered while the other orders are negligible in comparison to the first diffraction order. The magnification of the system at $z_2/2$ from the RMBDL is 1.

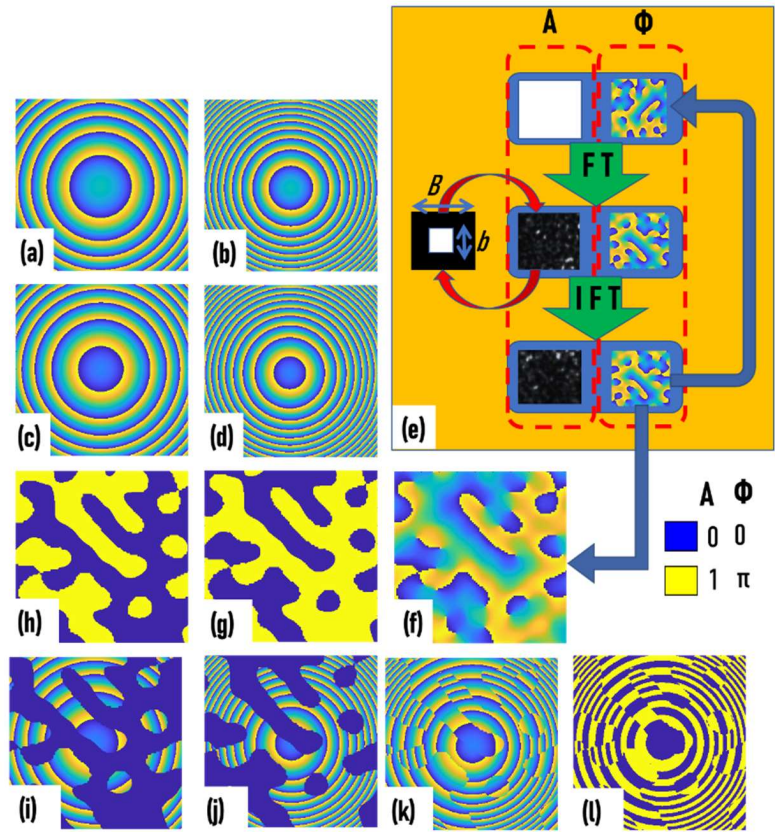

Fig. 2 Procedure for designing RMBDL. Phase images of the $FZL_1$ (a) before and (c) after aberration compensation. Phase images of $FZL_2$ (b) before and (d) after aberration compensation. (e) GSA for synthesizing a phase mask with a scattering ratio of $b/B$. (f) Phase mask synthesized from GSA. (g) Binary phase mask after binarization to two levels.

(h) Inverted image of the phase mask. Phase images of (i) FZL$_1$ and (j) FZL$_2$ after multiplying the binary phase mask. (k) Phase images of RMBDL (k) before and (l) after binarization to two levels.

## 2.2 Theoretical Analysis

A point object located at $(\bar{r}_o, z_1)$ emits light with an amplitude of $\sqrt{I_o}$ which reaches the RMBDL with a complex amplitude given by $C_1\sqrt{I_o}L(\bar{r}_o/z_1)Q(1/z_1)$, where $L(\bar{o}/z_1) = \exp[j2\pi(o_x x + o_y y)/(\lambda z_1)]$ is the linear phase factor and $\bar{r}_o = (x_o, y_o)$. It is assumed that the point object has a narrow spectral width and so a single wavelength was used for the following calculations. RMBDL modulates the incoming light and generates two waves in which one is focused at $z_2/2$ from the RMBDL while the other wave is collimated. The complex amplitude after the RMBDL is $C_1\sqrt{I_o}L(\bar{r}_o/u)Q(1/z_1)\phi_{RMBDL}$. For simplicity, let us consider only the first order diffraction patterns and assume the effect of glass substrate which cancels out the aberration compensation terms and therefore the complex amplitude after the RMBDL can be expressed as $C_1\sqrt{I_o}L(\bar{r}_o/z_1)Q(1/z_1)[\exp(-j4\pi r^2/\lambda z_2)(1-M) + \exp(-j\pi r^2/\lambda z_1)M]$. This results in two waves with equal intensities $C_2\sqrt{I_o}L(\bar{r}_o/z_1)[\exp(-j2\pi r^2/\lambda z_2)(1-M)]$ and $C_3\sqrt{I_o}L(\bar{r}_o/z_1)M$ where $C_2$ and $C_3$ are complex constants.

The rounding-off procedure converts equal number of pixels to 0 and 1 respectively resulting in a 50:50 splitting ratio. The complex amplitudes of the two waves at the image sensor located at $z_2$ from the RMBDL is given as $E_1 = C_2\sqrt{I_o}L(\bar{r}_o/z_1)[\exp(-j\pi r^2/\lambda z_1)(1-M)] \otimes Q(1/z_2)$ and $E_2 = C_3\sqrt{I_o}L(\bar{r}_o/z_1)M \otimes Q(1/z_2)$ where '$\otimes$' is a 2D convolutional operator. At the image sensor, the interference between the two waves can be written as $I_{PSH} = (E_1+E_2)^2$. If the point object is located on the optical axis, the linear phase factors can be neglected and the resulting pattern $I_{PSF}$ is an interference pattern between a plane wave and a spherical wave resulting in a circular fringe pattern. The presence of the random multiplexing matrices *M* and (1-*M*) introduces some multiplexing noises to the circular fringe pattern. As the illumination is incoherent, a complicated object may be considered as a collection of uncorrelated point objects given as $o(\overline{r_o}) = \sum_{i=1}^{M} a_i \delta(r - r_i)$. The object hologram $I_{OBJ}$ can be given as an addition of the circular fringe patterns corresponding to every object point. Therefore, $I_{OBJ} = I_{PSH} \otimes o(\overline{r_o})$ which can be reduced to $I_{OBJ} = \sum_{i=1}^{M} I_{PSH} \otimes a_i \delta(r - r_i)$. In the previous studies with three camera shots,[1,2,4-6] the image of the object is reconstructed by numerically propagating the complex holograms to the image plane of one of the two object waves. In the proposed method, the correlation relation between the object and the point object holograms has been utilized and the image of the object is reconstructed by a cross-correlation between the object and the point object holograms. The reconstructed image $I_R$ can be expressed as

$$I_R = \left| \mathcal{F}^{-1} \left\{ |\tilde{I}_{PSH}|^{\alpha} exp[i \, arg(\tilde{I}_{PSH})] |\tilde{I}_{OBJ}|^{\beta} exp[-i \, arg(\tilde{I}_{OBJ})] \right\} \right|, \qquad (2)$$

where the values of $\alpha$ and $\beta$ are tuned between -1 to +1 until a case with minimum entropy is obtained. The entropy is expressed as $S(\alpha, \beta) = -\sum\sum \phi(m,n) log[\phi(m,n)]$, where $\phi(m,n) = |C(m,n)|/\sum_M \sum_N |C(m,n)|$, $C(m,n)$ is the correlation distribution, and $(m,n)$ are the indexes of the correlation matrix. A comparison study of different types of filters for reconstructing the object from the hologram is presented in **Appendix A**.

# 3  Experiments

*3.1 Fabrication of RMBDL*

An Indium-Tin-Oxide (ITO) coated glass substrate with a thickness of 1.1 mm and index of refraction approximately 1.5 was ultrasonically cleaned in acetone and Iso-Propyl Alcohol (IPA) for 5 minutes. The substrate was dehydrated by baking it for 5 minutes at 180°C on a hotplate. After cooling to room temperature, tapes were pasted to the sides approximately 4 mm to mask ITO areas during spin coating for electrical contact during e-beam patterning. The substrate was spin coated with PMMA 950K A7 positive resist at 2000 RPM, ramp of 500 RPM/s for a duration of about a minute. The masking tapes were removed, and the substrate was baked at 180°C in a hot plate for 90 seconds. The substrate was loaded on to a substrate holder and the metal clip was attached to the masked area where ITO layer is not coated with resist. The design was fabricated on the resist using electron beam direct writing system RAITH 150$^{\text{TWO}}$ with 10kV acceleration voltage, 120 μm aperture, beam current of approximately ~3 nA, write field of 100 μm and a working distance of 10 mm. No stitching error could be observed using optical microscopy, demonstrating the writefield alignment done very well. The entire pattern was fabricated over a 6 hour period. The optical microscope images of the fabricated device are shown in Fig. 3. It is seen from the fabrication results that there is no stitching error.

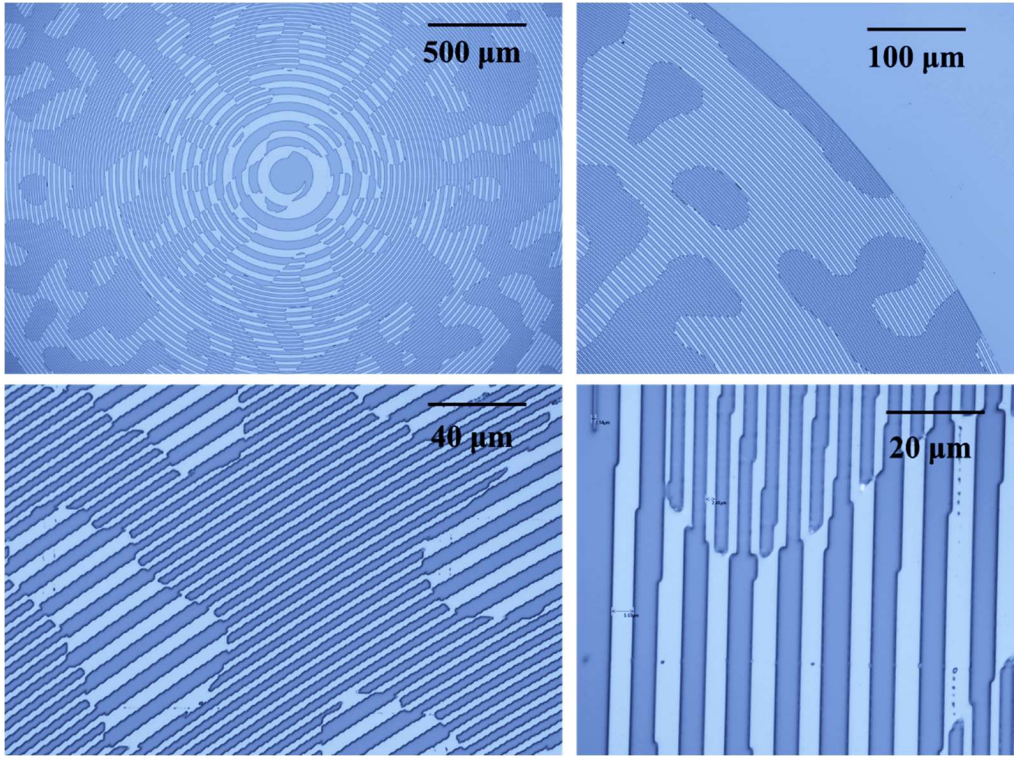

Fig. 3 Optical microscope images of the fabricated RMBDL.

*3.2 Imaging experiments*

An experimental set up was built as shown in Fig. 1 with an LED (M617L3, $\lambda_c$ = 617 nm, FWHM = 18 nm) which critically illuminates a pinhole with a diameter of 20 μm. The light diffracted from the pinhole is collected by the RMBDL located at a distance of 5 cm from it. The RMBDL splits the beam into two, focusing one at a distance of 5 cm and collimating the other beam. The PSH is recorded at a distance of 10 cm from the RMBDL by an image sensor (Thorlabs DCU223M, 1024 x 768 pixels, pixel size = 4.65 μm). In the first experiment, the location of the pinhole was shifted from $z_1$-3 cm to $z_1$+3 cm in steps of 5 mm and the corresponding PSHs are recorded. The lateral and axial resolutions of the system are given by $1.22\lambda z_1/D$ and $\sim 8\lambda(z_1/D)^2$ respectively which are

~7.5 µm and ~0.5 mm respectively. The images of the recorded PSHs are shown in Fig. 4. From the figures, it is seen that the best beam overlap condition was achieved only when the object is located at $z_1 = 5$ cm. The experiment was repeated with a step size of 1 mm and the corresponding PSHs are recorded and cross-correlated with the PSH recorded at $z_1=5$ cm with a phase-only filter. The plot of $I_R(x=0, y=0)$ with the location of the PSH is shown in Fig. 4(l).

The element 1 of Group 4 (16 lp/mm, grating period = 62.5 µm) of the United States Air Force (USAF) resolution target was mounted at $z_1 = 5$ cm and the image of the object was recorded at $z_2 = 5$ cm and the object hologram was recorded at $z_2 = 10$ cm. Since, the image of the object is recorded at $u = v = 5$ cm configuration, the magnification $M = 1$ and the features of the captured image matches with that of the object. The direct image is shown in Fig. 5(a) and the grating period was found to be 65.1 µm. The image of the hologram is shown in Fig. 5(b). The reconstruction results using Lucy-Richardson algorithm (250 iterations), Weiner filter, Fresnel back propagation and non-linear correlation ($\alpha = 0.2, \beta = 0.6$) are shown in Figs. 5(c)-(f) respectively. The complete reconstruction results of non-linear filter are given in **Appendix B**. From the results, it is seen that only non-linear filter and Lucy-Richardson algorithm reconstructed the image while Lucy-Richardson algorithm showed the highest SNR of all cases. However, Lucy-Richardson method resulted in a lossy reconstruction as the full object information was not obtained. The number '1' was not present in the reconstructed results which was recovered in the case of non-linear filter. In the following experiments, only Lucy-Richardson algorithm and non-linear filter were compared.

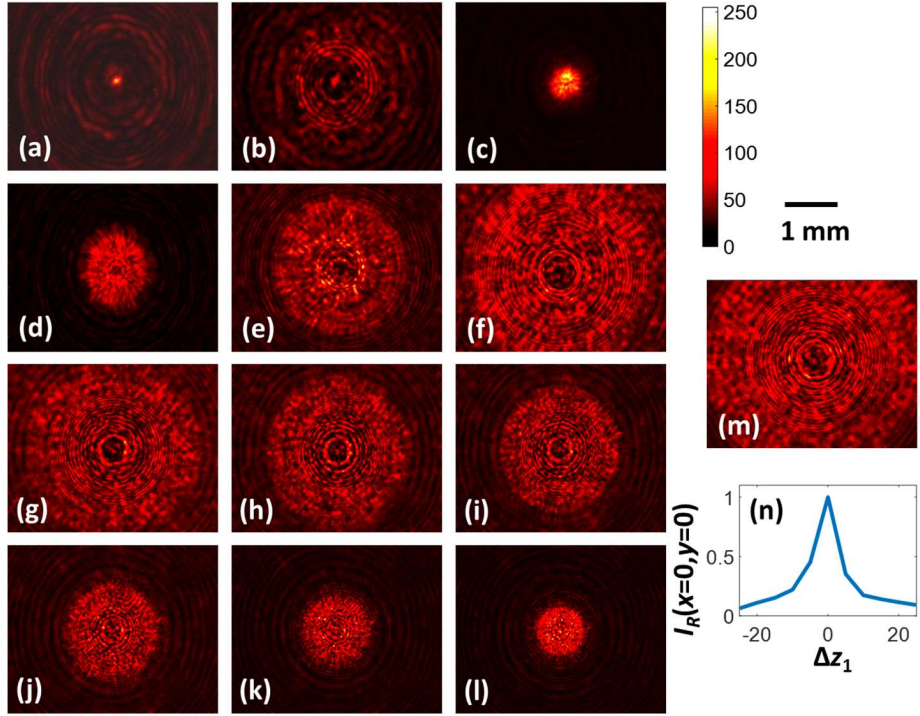

Fig. 4 Images of the PSHs recorded for $\Delta z_1$ = (a) -3 cm, (b) -2.5 cm, (c) -2 cm, (d) -1.5 cm, (e) -1 cm, (f) -0.5 cm, (g) 0.5 cm, (h) 1 cm, (i) 1.5 cm, (j) 2 cm, (k) 2.5 cm, (l) 3 cm and (m) 0 cm. (n) Plot of the variation of $I_R(x=0,y=0)$ as a function of $\Delta z_1$.

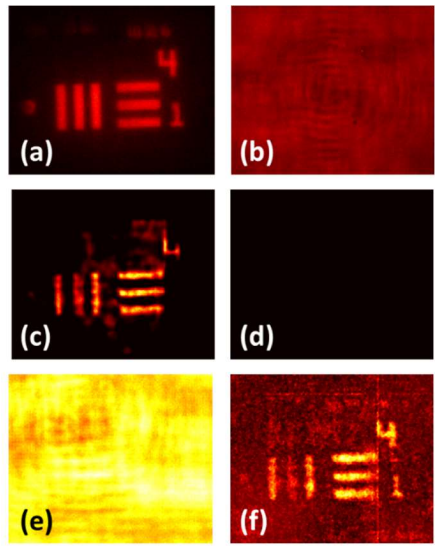

Fig. 5 (a) Direct imaging result of the USAF object recorded at $z_2$ = 5 cm. (b) FINCH hologram of the USAF object. Reconstruction results using (c) Lucy-Richardson algorithm (250 iterations), (d) Weiner filter, (e) Fresnel back propagation and (f) non-linear filter ($\alpha$ = 0.2, $\beta$ = 0.6).

As the spacing between the object plane and the RMBDL is only 5 cm, a two-channel experiment is difficult to perform with two planes. Therefore, the three-dimensional image reconstruction has been demonstrated using a synthetic hologram of a two plane object. The synthetic hologram is generated based on the principles of incoherent imaging where there is only an intensity addition and not interference. In all the previous studies,[1,2,4-6] an interference between the two objects at different axial locations was avoided by placing the two objects laterally separated from one another. A second object, element '14' from the National Bureau of Standards (NBS) mask was mounted at different axial locations $z_1$-1 cm to $z_1$+1 cm in steps of 5 mm. The images of the NBS holograms for the object locations $z_1$-1 cm to $z_1$+1 cm in steps of 5 mm are shown in Fig. 7. The synthetic holograms were obtained by adding the NBS holograms at different axial locations to the USAF hologram recorded at $z_1$ = 5 cm. The images of the synthetic holograms for different locations of the NBS object are shown in Fig. 6. The reconstruction results using the PSH at $z_1$=5 cm and $z_1$ corresponding to the locations of the NBS object using both Lucy-Richardson algorithm and non-linear filter are shown in Fig. 6. The direct imaging of the NBS object is shown as an inset in Fig. 6. The experiments at the resolution limit of the system are given in **Appendix C** and the experiments with a stained biological sample are discussed in **Appendix D**.

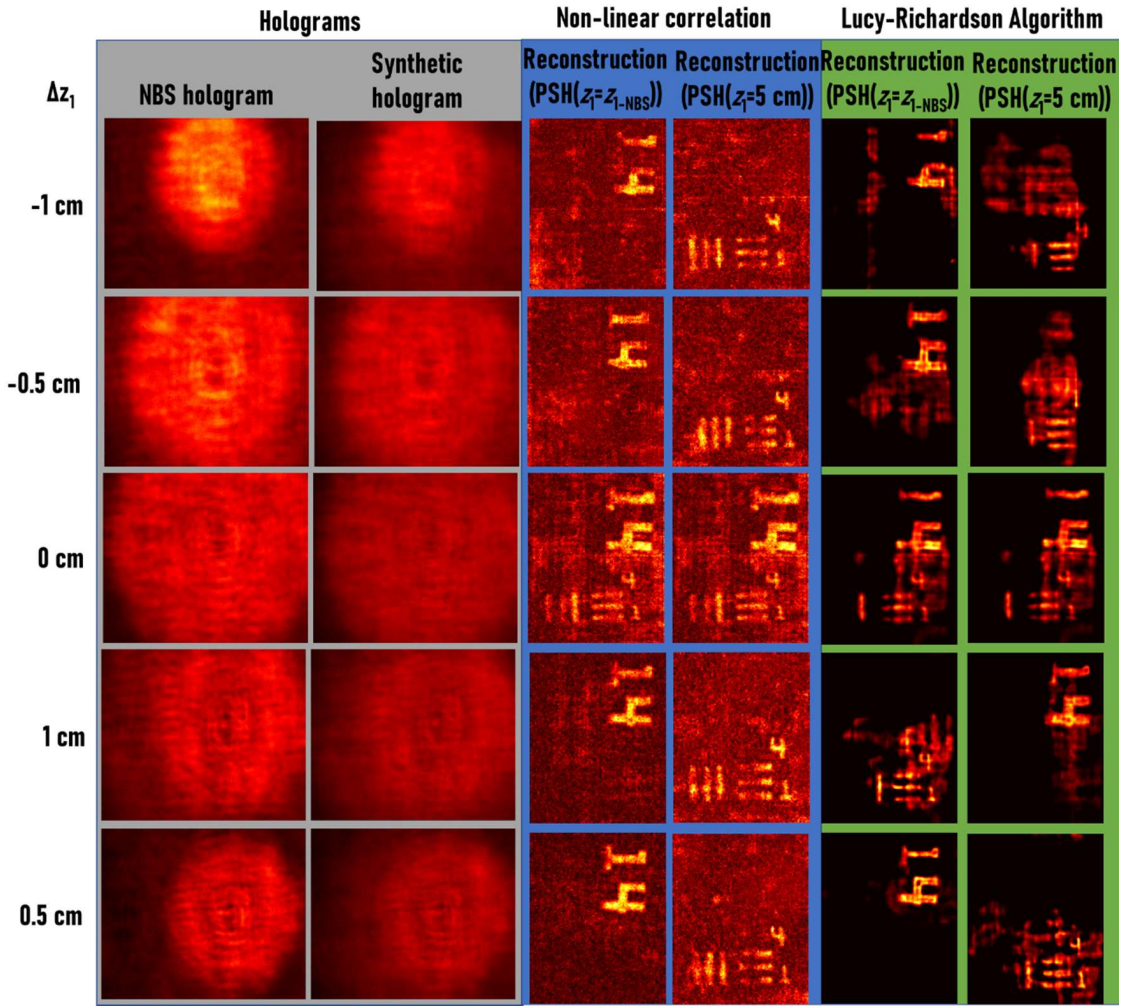

Fig. 6 Three-dimensional reconstruction results for synthetic holograms generated from the holograms of USAF and NBS object are presented. The thickness of the synthetic hologram was increased from -1 cm to 1 cm in steps of 5 mm and the holograms were reconstructed using the PSHs recorded at different locations using Lucy-Richardson algorithm and non-linear filter.

## 4 Summary and Conclusion

The FINCH technique has been demonstrated using a single diffractive optical element RMBDL in a compact optical configuration with a single camera shot. In the previous studies, FINCH has been demonstrated using at least three camera shots with phase shift and therefore required an active device such as a spatial light modulator. Recent developments on achieving a single camera shot imaging with FINCH has shown only little progress with a sacrifice of either the field of view or reconstruction quality or both.[8,10,11] Besides, none of the above studies demonstrated the super resolution capability of FINCH. In this study, FINCH has been converted from multiple elements optical system to a single element optical system (RMBDL) with a compact optical configuration. A modified approach based on manual PSH training and cross-correlation principle was adapted for the first time for FINCH. Finally, various reconstruction techniques have been compared to reconstruct object with the highest SNR. Two techniques namely Lucy-Richardson algorithm and non-linear filter seem promising for single camera shot FINCH. Lucy-Richardson algorithm has a higher SNR in comparison to the non-linear filter, however seems lossy as some of the object information was lost during reconstruction. On the other hand, the non-linear filter reconstructed the object information completely, while has a lower SNR due to the background noise. For both cases, reconstruction results for small objects was found to be better in comparison to larger objects. The current configuration of FINCH produced a higher lateral resolving power but equivalent axial resolving power to direct imaging technique. Further studies are necessary in order to improve the SNR of reconstruction and demonstrate the maximum super resolution capability of FINCH. We believe that the proposed optical configuration and reconstruction mechanism will improve the latest model of FINCH.[22]

**Appendix A: Comparison of different decorrelation techniques**

FINCH with polarization multiplexing scheme where all the SLM pixels were utilized for the generation of the two beams is compared with random multiplexing scheme with different scattering ratio masks using the recently developed non-linear correlation technique. The above FINCH cases, were also studied using well-established decorrelation methods such as Lucy-Richardson method, Weiner Filter and Fresnel back propagation. The design values described in section 2(a) is used for simulation and Swinburne University's emblem is used as a test object.

The images of the point object holograms and object holograms for the above FINCH cases are shown in Fig. 7. The reconstruction results using Lucy-Richardson method, Weiner Filter, Fresnel back propagation and non-linear filter with minimum entropy are shown in Fig. 7. From the reconstruction results, it is seen that Fresnel propagation means of reconstruction is not successful due to the presence of twin image and bias terms. Non-linear reconstruction produced the image but is seen noisy. Lucy-Richardson algorithm needed about 200 iterations with the 'deconvlucy' function of MATLAB to reconstruct the image, but the reconstructed image was blurred and the time consumed was 120 seconds (Intel core i5-8250U CPU 1.6 GHz, 1.8 GHz, 8 Gigabytes RAM). Therefore, Lucy-Richardson algorithm may not be ideal for real-time three-dimensional imaging but can be used to record events for later analysis. Another observation in the results of Lucy-Richardson algorithm is that with a decrease in the scattering ratio, the blur in the reconstructed results appear to increase. Weiner filter produced results identical to that of the image. However, previous studies on Weiner filter under non-ideal noisy configurations showed a lower SNR.[21] From the study, it is seen that the scattering ratio does not have much effect on the reconstruction results. A higher scattering ratio increases the computational data and the size of the CAD file. The design was generated with a scattering ratio $\sigma = 0.1$ and a diameter of 5 mm as a bitmap file and converted into GDSII format using the trial version of LinkCAD software. The file size was approximately 65 Megabytes.

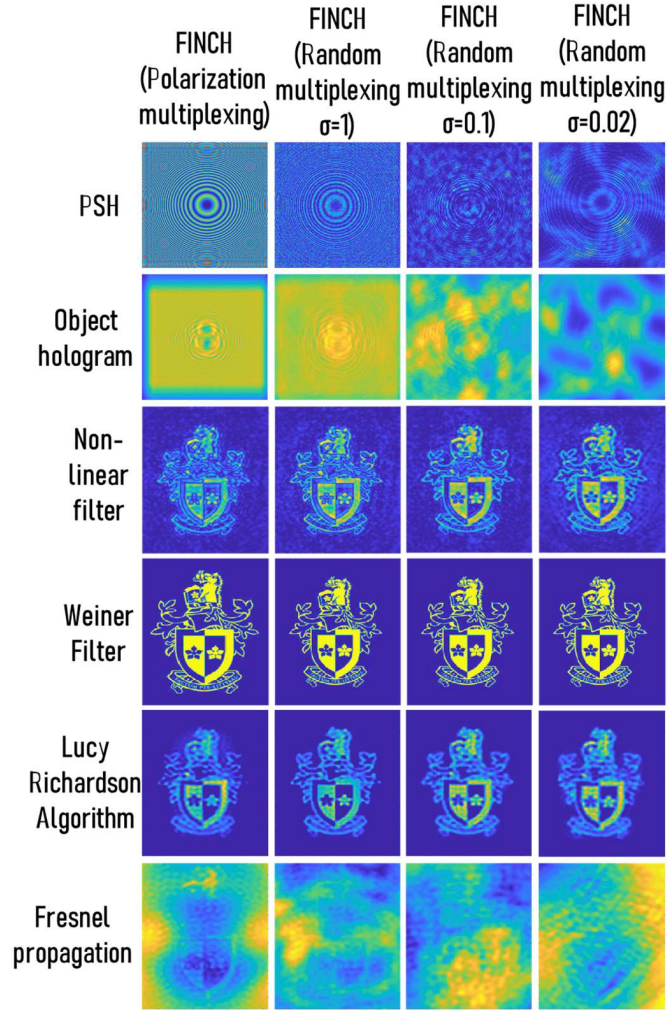

Fig. 7 Reconstruction results of FINCH in different configurations using different decorrelation techniques such as non-linear filter, Weiner filter, Lucy-Richardson filter (200 iterations) and Fresnel propagation.

**Appendix B: Reconstruction by non-linear filter**

The values of $\alpha$ and $\beta$ were varied in steps of 0.2 from -1 to +1 and a low pass filter and median filter was implemented simultaneously and the entropy was calculated for each case. The reconstruction results of the USAF object is shown in Fig. 8. The reconstruction results for negative values of $\alpha$ did not produce any result.

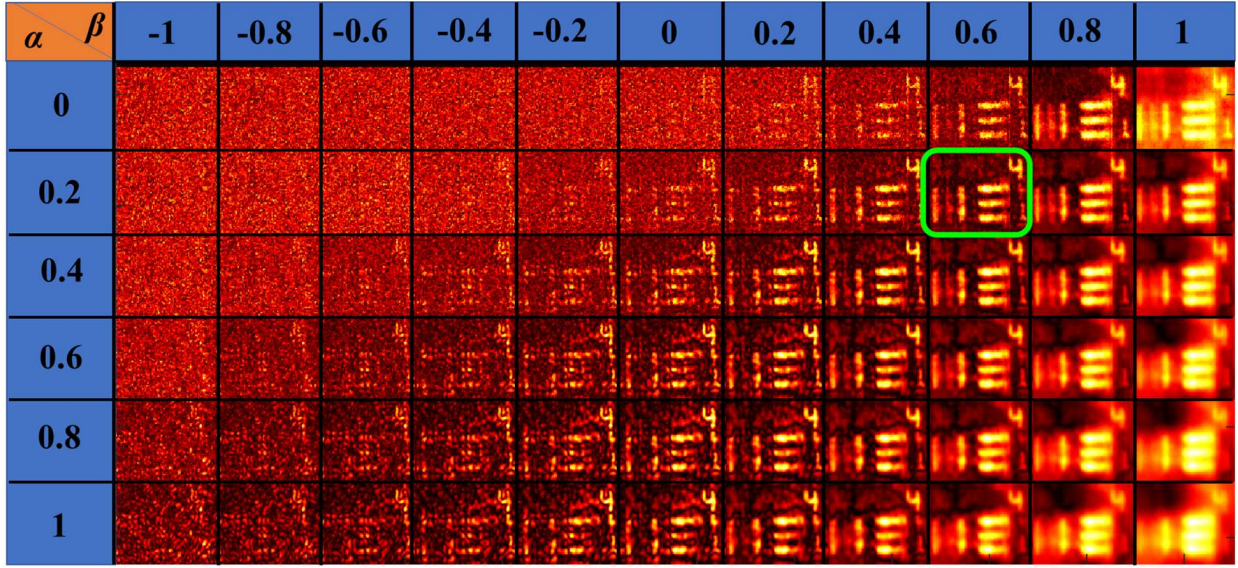

Fig. 8 Reconstruction results using the non-linear filter for different values of α and β. The result with the lowest entropy is indicated by green box.

**Appendix C**

The experiment was repeated at the resolution limit of the imaging system. A USAF object Group 5, Elements 5 (50.8 lp/mm) and 6 (57 lp/mm) were imaged using FINCH and the object and point object holograms are shown in Fig. 9(a) and 9(b) respectively. The objects were imaged using direct imaging system and the resolution limit was verified. The element 5 of Group 5 was barely resolved while the element 6 was completely unresolved where the three grating lines could not be perceived as three distinct lines as shown in Fig. 10(a). The normalized average visibility plot of the lines are shown in Fig. 10(b) and the average visibility was found to be 0.06. In the case of non-linear correlation, the three grating lines of Element 5 are well-resolved while in Element 6 the three grating lines can be perceived distinctly as shown in Fig. 10(c) unlike the direct imaging case. The normalized average visibility plot is shown in Fig. 10(d) with a visibility value of 0.7. Lucy-Richardson algorithm was used with 150 iteration and the result and the normalized average visibility plot are shown in Figs. 10(e) and 10(f) respectively. The visibility value was 0.7. When Lucy-Richardson algorithm was used with 200 iterations the results improved as shown in Figs.

10(g) and 10(h). The visibility value was found to be 0.77. As seen earlier, the non-linear correlation was noisy while Lucy-Richardson is lossy. However, in both cases, the resolution enhancement as expected in FINCH is clearly visible. Therefore, this is the first study where the enhanced resolution is demonstrated with a single camera shot. However, further studies are necessary to understand the conditions required to achieve the maximum resolution.

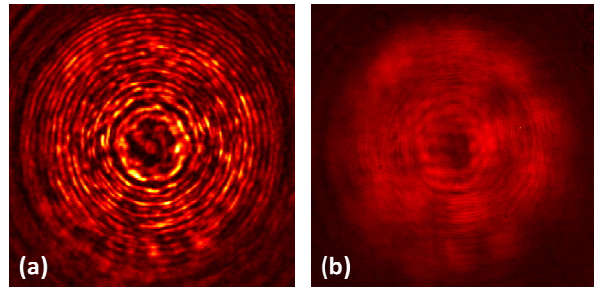

Fig. 9 Images of the recorded (a) PSH and (b) object hologram of elements 5 and 6 of group 5.

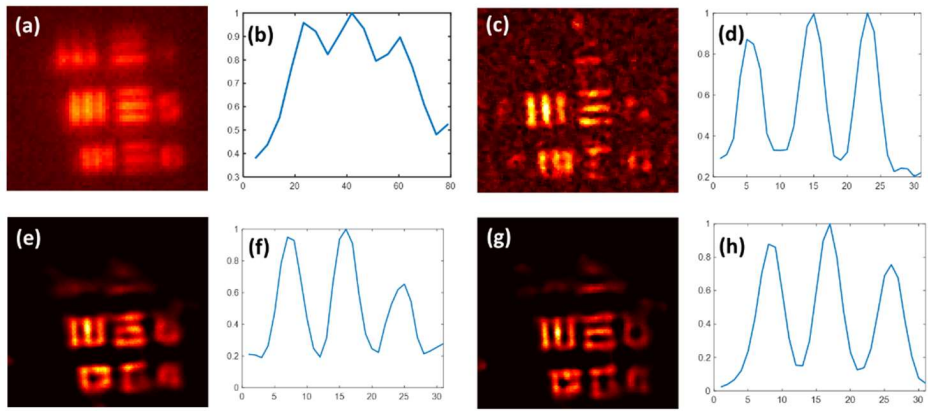

Fig. 10 (a) Direct imaging result of elements 5 and 6 of group 5 and (b) normalized average visibility plot of the gratings. Reconstruction result from (c) non-linear filter and Lucy-Richardson algorithm (e) 150 iterations and (f) 200 iterations. Normalized average visibility plot of the gratings for (d) non-linear filter, Lucy-Richardson algorithm (f) 150 iterations and (g) 200 iterations.

**Appendix D**

The experiment was repeated using a biological sample. The sample is a histological thin-section, approximately 0.25 micron in thickness, of the dragonfly larvae wing. The thin section represents the cross-section of the body of the wing showing the morphology of the larval wing at an early stage. The wing was stained using heavy metals then embedded in epoxy resin. Thin-sections were prepared using an ultramicrotome. Thin sections are mounted on a glass slide and stained with Toluidine blue. The image of the sample captured using a regular Nikon microscope is shown in Fig. 11(a). The same image recorded using direct imaging method with incoherent illumination with red wavelength using the limited NA of RMBDL is shown in Fig. 11(b). The images of the object hologram and PSH are shown in Figs 12(a) and 12(b), the reconstructed images using non-linear filter and Lucy-Richardson algorithm are shown in Fig. 12(c) and 12(d) respectively. Comparing Fig. 11(b) with Figs. 12(c) and 12(d) shows that FINCH has a higher lateral resolution than the direct imaging and additional features are visible in FINCH unlike direct imaging. The reconstruction results of non-linear filter using the PSF library recorded between -4 mm and 3 mm depths from $z_1 = 5$ cm are shown in Fig. 13. The different layers of the specimen are focused when reconstructed with PSHs recorded at those locations.

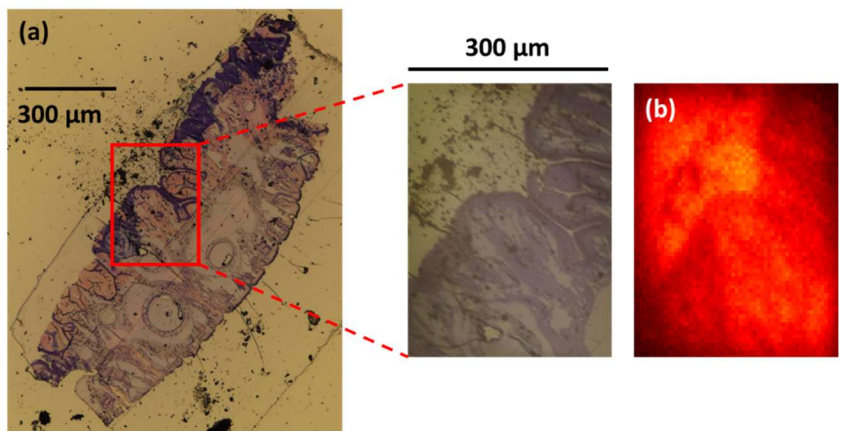

Fig. 11 (a) Microscope image of the stained biological sample. The area within the rectangular red box was used for the experiments while the rest of the area was masked with a black tape. (b) The direct imaging result with the RMBDL recorded at $z_2$ = 5 cm from the RMBDL.

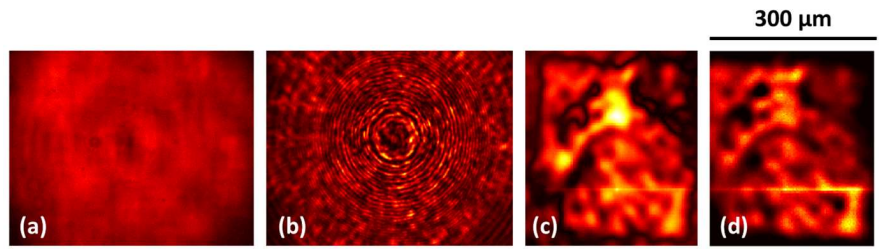

Fig. 12 (a) Object hologram, (b) PSH, reconstruction results using (c) non-linear filter and (d) Lucy-Richardson algorithm.

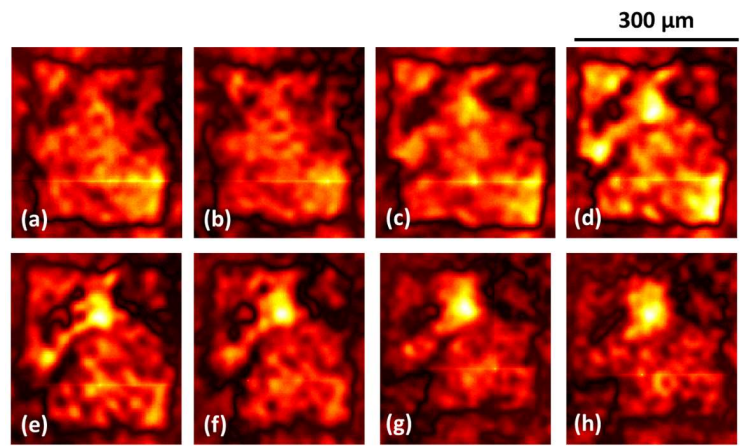

Fig. 13 (a)-(h) Reconstruction results using non-linear filter with PSHs recorded at distances $\Delta z_1$ = -4 mm to 3 mm.

*Disclosures*

The authors do not have any conflict of interests.

*Acknowledgments*

NATO grant No. SPS-985048 is acknowledged for funding.

*References*

1. J. Rosen, and G. Brooker, "Digital spatially incoherent Fresnel holography," *Opt. Lett*. **32**(8), 912-914 (2007).

2. J. Rosen, and G. Brooker, "Non-Scanning Motionless Fluorescence Three-Dimensional Holographic Microscopy," *Nat. Photon*. **2**, 190-195 (2008).

3. T-C. Poon, "Optical Scanning Holography - A Review of Recent Progress," *J. Opt. Soc. Korea* **13**(4), 406-415 (2009).

4. G. Brooker, N. Siegel, V. Wang and J. Rosen, "Optimal resolution in Fresnel incoherent correlation holographic fluorescence microscopy," *Opt. Express* **19**(6), 5047-5062 (2011).

5. J. Rosen, N. Siegel, and G. Brooker, "Theoretical and experimental demonstration of resolution beyond the Rayleigh limit by FINCH fluorescence microscopic imaging," *Opt. Express* **19**(27), 26249-26268 (2011).

6. B. Katz, J. Rosen, R. Kelner and G. Brooker, "Enhanced resolution and throughput of Fresnel incoherent correlation holography (FINCH) using dual diffractive lenses on a spatial light modulator (SLM)," *Opt. Express* **20**(8), 9109-9121(2012).

7. R. Kelner and J. Rosen, "Spatially incoherent single channel digital Fourier holography" *Opt. Lett.* **37**(17), 3723–3725 (2012).

8. T. Tahara, T. Kanno, Y. Arai, and T. Ozawa, "Single-shot phase-shifting incoherent digital holography," *J. Opt*. **19**(6), 065705 (2017).

9. T. Nobukawa, T. Muroi, Y. Katano, N. Kinoshita, and N. Ishii, "Single-shot phase-shifting incoherent digital holography with multiplexed checkerboard phase gratings," *Opt. Lett*. **43**(8), 1698-1701 (2018).

10. X. Quan, O. Matoba, and Y. Awatsuji, "Single-shot incoherent digital holography using a dual-focusing lens with diffraction gratings," *Opt. Lett*. **42**(3), 383-386 (2017).
10

**Caption List**

**Fig. 1** Optical configuration of FINCH with a RMBDL and with a modified image reconstruction.

**Fig. 2** Fig. 2 Procedure for designing RMBDL. Phase images of the $FZL_1$ (a) before and (c) after aberration compensation. Phase images of $FZL_2$ (b) before and (d) after aberration compensation. (e) GSA for synthesizing a phase mask with a scattering ratio of $b/B$. (f) Phase mask synthesized from GSA. (g) Binary phase mask after binarization to two levels. (h) Inverted image of the phase mask. Phase images of (i) $FZL_1$ and (j) $FZL_2$ after multiplying the binary phase mask. (k) Phase images of RMBDL (k) before and (l) after binarization to two levels.

**Fig. 3** Optical microscope images of the fabricated RMBDL.

**Fig. 4** Images of the PSHs recorded for $\Delta z_1$ = (a) -3 cm, (b) -2.5 cm, (c) -2 cm, (d) -1.5 cm, (e) -1 cm, (f) -0.5 cm, (g) 0.5 cm, (h) 1 cm, (i) 1.5 cm, (j) 2 cm, (k) 2.5 cm, (l) 3 cm and (m) 0 cm. (n) Plot of the variation of $I_R(x=0,y=0)$ as a function of $\Delta z_1$.

**Fig. 5** (a) Direct imaging result of the USAF object recorded at $z_2$ = 5 cm. (b) FINCH hologram of the USAF object. Reconstruction results using (c) Lucy-Richardson algorithm (250 iterations), (d) Weiner filter, (e) Fresnel back propagation and (f) non-linear filter ($\alpha$ = 0.2, $\beta$ = 0.6).

**Fig. 6** Three-dimensional reconstruction results for synthetic holograms generated from the holograms of USAF and NBS object are presented. The thickness of the synthetic hologram was increased from -1 cm to 1 cm in steps of 5 mm and the holograms were reconstructed using the PSHs recorded at different locations using Lucy-Richardson algorithm and non-linear filter.

**Fig. 7** Reconstruction results of FINCH in different configurations using different decorrelation techniques such as non-linear filter, Weiner filter, Lucy-Richardson filter (200 iterations) and Fresnel propagation.

**Fig. 8** Reconstruction results using the non-linear filter for different values of $\alpha$ and $\beta$. The result with the lowest entropy is indicated by green box.

**Fig. 9** Images of the recorded (a) PSH and (b) object hologram of elements 5 and 6 of group 5.

**Fig. 10** (a) Direct imaging result of elements 5 and 6 of group 5 and (b) normalized average visibility plot of the gratings. Reconstruction result from (c) non-linear filter and Lucy-Richardson algorithm (e) 150 iterations and (f) 200 iterations. Normalized average visibility plot of the gratings for (d) non-linear filter, Lucy-Richardson algorithm (f) 150 iterations and (g) 200 iterations.

**Fig. 11** (a) Microscope image of the biological sample. The area within the rectangular red box was used for the experiments while the rest of the area was masked with a black tape. (b) The direct imaging result with the RMBDL recorded at $z_2 = 5$ cm from the RMBDL.

**Fig. 12** (a) Object hologram, (b) PSH, reconstruction results using (c) non-linear filter and (d) Lucy-Richardson algorithm.

**Fig. 13** (a)-(d) Reconstruction results using non-linear filter with PSHs recorded at distances $\Delta z_1$ = -4 mm to 3 mm.